\documentclass[twocolumn,
                showpacs,
                nofootinbib,
                nobibnotes,
                aps,
                superscriptaddress,
                amssymb,
                amsmath,
                floatfix,
                reprint,
                prx, 
                notitlepage,
                showkeys,
                10pt,
		]{revtex4-2}

\usepackage{amssymb}
\usepackage{amsmath}
\usepackage{cprotect}
\usepackage{graphicx}
\usepackage{dcolumn}
\usepackage{url}
\usepackage[colorlinks=true,breaklinks=true,allcolors=blue]{hyperref}
\usepackage{tikz}
\usepackage{dsfont}
\usepackage{epsfig}
\usepackage{color}
\usepackage[ruled]{algorithm2e}
\usepackage{bbold}
\usepackage{glossaries}
\usepackage{xcolor}
\usepackage{tikz}
\usepackage{bm}
\usepackage{hhline}
\usepackage{spverbatim}
\usepackage{siunitx}
\makeatletter
\let\cat@comma@active\@empty
\makeatother

\begin{document}
\title{Enhancing variational Monte Carlo using a programmable quantum simulator}

\author{M. Schuyler Moss}
\email{msmoss@uwaterloo.ca}
\affiliation{Department of Physics and Astronomy, University of Waterloo, Ontario, N2L 3G1, Canada}
\affiliation{Perimeter Institute for Theoretical Physics, Waterloo, Ontario, N2L 2Y5, Canada}

\author{Sepehr Ebadi}
\affiliation{Department of Physics, Harvard University, Cambridge, MA 02138, USA}

\author{Tout T. Wang}
\affiliation{Department of Physics, Harvard University, Cambridge, MA 02138, USA}

\author{Giulia Semeghini}
\affiliation{Department of Physics, Harvard University, Cambridge, MA 02138, USA}

\author{Annabelle Bohrdt}
\affiliation{Department of Physics, Harvard University, Cambridge, MA 02138, USA}
\affiliation{ITAMP, Harvard-Smithsonian Center for Astrophysics, Cambridge, MA, USA}
\affiliation{Institut für Theoretische Physik, Universität Regensburg, D-93035 Regensburg, Germany}
\affiliation{Munich Center for Quantum Science and Technology (MCQST), D-80799 M\"unchen, Germany}

\author{Mikhail D. Lukin}
\affiliation{Department of Physics, Harvard University, Cambridge, MA 02138, USA}

\author{Roger G. Melko}
\affiliation{Department of Physics and Astronomy, University of Waterloo, Ontario, N2L 3G1, Canada}
\affiliation{Perimeter Institute for Theoretical Physics, Waterloo, Ontario, N2L 2Y5, Canada}

\date{August 4, 2023} 

\begin{abstract}
Programmable quantum simulators based on Rydberg atom arrays are a fast-emerging quantum platform, bringing together long coherence times, high-fidelity operations, and large numbers of interacting qubits deterministically arranged in flexible geometries.  Today's Rydberg array devices are demonstrating their utility as quantum simulators for studying phases and phase transitions in quantum matter. In this paper, we show that unprocessed and imperfect experimental projective measurement data can be used to enhance {\it in silico} simulations of quantum matter, by improving the performance of variational Monte Carlo simulations.  As an example, we focus on data spanning the disordered-to-checkerboard transition in a $16 \times 16$ square lattice array [S. Ebadi {\it et al.} Nature  595, 227 (2021)] and employ data-enhanced variational Monte Carlo to train powerful autoregressive wavefunction ans\"atze based on recurrent neural networks (RNNs). We observe universal improvements in the convergence times of our simulations with this hybrid training scheme. Notably, we also find that pre-training with experimental data enables relatively simple RNN ans\"atze to accurately capture phases of matter that are not learned with a purely variational training approach. Our work highlights the promise of hybrid quantum--classical approaches for large-scale simulation of quantum many-body systems, combining autoregressive language models with experimental data from existing quantum devices.
\end{abstract}

\maketitle

%
\section{Introduction} 
%
A central challenge for the current generation of quantum computers and simulators is determining whether their outputs can provide value for problems of practical interest.
This is particularly urgent for devices that have sufficiently large numbers of qubits, long coherence times, and high connectivity, such that their outputs cannot be simulated exactly using classical methods~\cite{googleadvantage,CASadvantage,xanaduadvantage,lukinadvantage}.
Programmable quantum devices made of arrays of Rydberg atoms are no exception, with rapid scaling progress being achieved by several groups in academia and industry~\cite{atom-by-atom,51qubits,100qubits,Henriet2020quantumcomputing,100qubits-afm,200qubits,hardware-eff-fault-tol,Huber:22,QuEra-optimization}.
The large number of qubits in the latest experimental Rydberg arrays have allowed these systems to flourish as quantum simulators of quantum matter, with numerous impressive demonstrations preparing exotic phases and studying their phase transitions~\cite{100qubits-afm,symprot,ebadi2021quantum,ryd-spinliq} and dynamics~\cite{scars,51qubits}.
It is intriguing to inquire if such devices can be used to accelerate solutions to practical simulation
problems despite existing limitations to system size, simulation time, decoherence and data acquisition rate.

\begin{figure*}[!]
    \centering
    \includegraphics{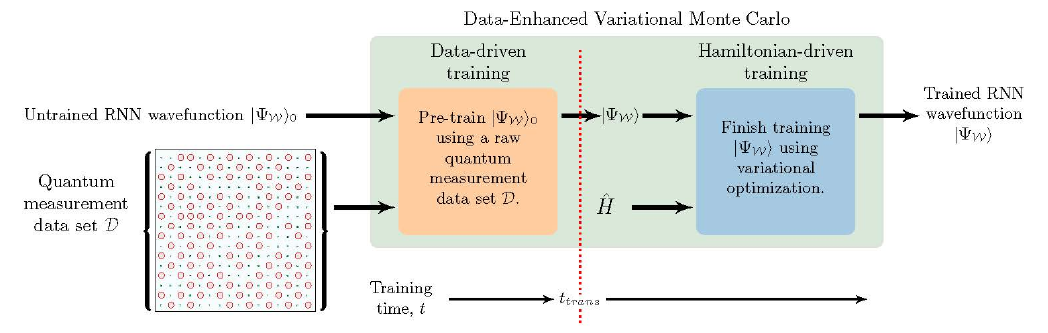}
    \vspace*{-3mm}
    \caption{
    A schematic outlining the data- then Hamiltonian-driven training procedure we refer to as data-enhanced variational Monte Carlo. First a randomly initialized recurrent neural network (RNN) wavefunction $\vert \Psi_{\mathcal{W}}\rangle_0$ is trained on quantum measurement data. In this case, our quantum measurement data set $\mathcal{D}$ contains fluorescent images of projective measurements made on Rydberg atom arrays, such as the one shown in brackets. The fluorescent imaging technique only captures atoms in the electronic ground state $\vert g \rangle$, the small green points, whereas the atoms in the excited state $\vert r \rangle$ are detected as a loss and marked with red circles. After $t_{\text{trans}}$ data-driven training steps, the pre-trained RNN wavefunction $\vert \Psi_{\mathcal{W}}\rangle$ is trained variationally, making use of the presumed many-body Rydberg Hamiltonian $\hat{H}$. The red dotted line marks the transition from data-driven training to Hamiltonian-driven training. The final, trained RNN wavefunction $\vert \Psi_{\mathcal{W}}\rangle$ gives an approximation to the ground state wavefunction of $\hat{H}$.
    }
    \label{fig:schematic}
\end{figure*}

Even for today's best quantum computing platforms, determining properties of the quantum state is fundamentally hampered by the intractability of traditional tomography.  Namely, since the state space grows exponentially, a full tomographic reconstruction will always be data-limited, making it impossible for any platform with a large number of qubits.
One useful strategy that has recently emerged for extracting value from a limited amount of quantum measurement data is generative modelling~\cite{torlai2018neural}.
Generative neural networks have many desirable properties that make them especially suitable as ansatz quantum states for learning tasks in quantum many-body physics. 
For example, they are flexible architectures that can be readily extended to represent higher dimensional systems, they can be trained with efficient heuristics, and improvements can be systematically explored through hyper-parameter tuning.
Models such as restricted Boltzmann machines (RBMs)~\cite{RBM_review,lange2022adaptive} and recurrent neural networks (RNNs)~\cite{Juan19generative,czischek2022data,iouchtchenko2023neural} have demonstrated successful wavefunction reconstruction, although the quality of the reconstructed state is still highly dependent on the number of measurement samples available and the quality of the data.
One recent RBM reconstruction of the ground state of a chain of only $N = 8$ Rydberg atoms used $\mathcal{O}(10^3)$ samples, requiring some degree of error mitigation applied to the experimental data~\cite{torlai2019integrating}.
As Rydberg array sizes continue to grow, the amount of data needed for an accurate reconstruction is also expected to grow, although a study of the sample complexity required for accurate wavefunction reconstruction has not been performed for Rydberg array data to date.  
As an example, information from $\mathcal{O}(10^3)$ measurements would not be sufficient to reconstruct the full quantum state for system sizes of $N \gg 8$ at least in the described manner; however, this data can still provide value in the training of generative models.

In this work, we demonstrate how limited and imperfect experimental data from a Rydberg atom array can be used without pre-processing or error mitigation to help in the variational training of generative models, specifically recurrent neural networks (RNNs), to accurately capture ground state wavefunctions at many points across a phase transition. 
We use a \emph{hybrid} training scheme introduced as data-enhanced variational Monte Carlo~\cite{czischek2022data} (or, neural error mitigation \cite{bennewitz2022neural}), that is both data-driven and Hamiltonian-driven (i.e.~variational). Fig.~\ref{fig:schematic} shows a schematic outlining this training procedure.
We find pre-training our RNN ans\"atze with the experimental data results in a robust improvement in the time-to-solution for the subsequent variational optimization.  
In addition to these speed-ups in convergence time, in some cases, data-enhancement improves the accuracy of the quantum state learned by the model. 
Once trained, an RNN wavefunction acts as a generative model, generating new wavefunction configurations with tractable autoregressive sampling, allowing for the calculation of many observables of interest, including those that cannot be calculated directly from projective measurement data. We provide results for estimators of the energy, the order parameter characterizing the phase transition, and an off-diagonal operator, confirming that our trained models capture characteristic features of the target ground states.
Our results indicate that generative autoregressive models designed for use as parameterized wavefunctions, such as RNNs, are well-suited to take advantage of raw unprocessed measurement data from today's quantum devices, enabling a potentially powerful hybrid approach for quantum simulators.

%
\section{Programmable Rydberg Atom Arrays}
%
As a concrete example, we focus on experimental work by Ebadi {\it et al.} \cite{ebadi2021quantum}, where two dimensional (2D) defect-free arrays containing hundreds of $^{87}$Rb atoms were studied. These arrays are assembled into an $L\times L$ square lattice using a combination of static and movable optical tweezers. The qubits in the array are encoded in the ground $\vert g \rangle$ and the highly excited Rydberg state $\vert r\rangle$, and the coherent evolution of the array is governed by the Rydberg Hamiltonian which is parameterized by the Rabi frequency $\Omega(t)$, detuning $\delta(t)$, and long-range interactions $V_{ij}$:
\begin{equation}
    \hat{H} = \frac{\Omega}{2}\sum_{i=1}^{N}\hat{\sigma}^{x}_{i} - \delta \sum_{i=1}^{N}\hat{n}_i + \sum_{i,j}V_{ij}\hat{n}_i\hat{n}_j,
\label{Hamiltonian}
\end{equation}
where $N = L^2$ is the number of qubits, $\hat{\sigma}^{x}_{i} = \vert g \rangle_i \langle r\vert_i + \vert r \rangle_i \langle g\vert_i $, 
$\hat{n}_i = \vert r\rangle_i \langle r \vert_i$, and
$V_{ij} = \Omega R_b^6/\vert \mathbf{r}_i - \mathbf{r}_j\vert^6$.
The Rydberg blockade mechanism~\cite{lukin2001dipole}, which penalizes the simultaneous excitation of atoms within a {\it blockade} radius $R_b \equiv (V_0/\Omega)^{1/6}$, can effectively be controlled by adjusting the lattice spacing $a$ of the atom array. 
Thus, by tuning the various parameters and programming the geometry of the array, a rich phase diagram of quantum states can be accessed using these devices.

After evolution under this Hamiltonian, the state of each atomic qubit is read out with a projective measurement where atoms in $| g \rangle$ are imaged, while atoms in $| r \rangle$ are ejected from the system.
This measurement procedure yields a set of binary data $\boldsymbol\sigma = 
\{n_i\}_{i=1}^{N}$, where $n_i = 0,1$, that represents one projective measurement of the many-body quantum state, which we refer to as a ``shot''.  
Because the Hamiltonian describing this system can be made stoquastic in the Rydberg occupation basis, projective measurements taken in this basis provide complete information about the positive, real-valued amplitudes of the ground state wavefunction.

Making use of this protocol, Ebadi {\it et al.} realized and explored a number of phases and phase transitions detected on a square $16 \times 16$ Rydberg atom array~\cite{ebadi2021quantum}. 
Here, we study the most well understood of these, the disordered and $Z_2$ checkerboard phases, including the quantum phase transition between them that lies in the (2+1)$D$ Wilson-Fisher (WF) universality class.
For a given $R_b/a$, a state of interest is prepared by linearly increasing $\delta$ at a given sweep rate $s$ and at fixed $\Omega$, from an initial large negative value with all atoms in $| g \rangle$, and stopping at the desired value of $\delta$.
We consider the slowest sweep rate, $s=15$ MHz/$\mu$s, for which we have the largest number of measurements (1000 per $\delta$) and the state preparation is most adiabatic. 
Each time a state is prepared, a single projective measurement is captured, requiring repeated state preparations and measurements for each final detuning value.
This raw experimental measurement data obtained for a range of detuning values that spans the disordered-to-checkerboard transition is used in all numerical experiments herein. 
For a more detailed list of the experimental parameters, see Appendix~\ref{app:params}.

For these experiments, measurements are of high quality, with readout fidelities of $99\%$ for both $\vert g \rangle$ and $\vert r \rangle$.
However, it is worth emphasizing that the states prepared on these devices are \emph{low energy states} and not necessarily the many-body ground states. 
As a result, measurement shots in the data set, especially those in the ordered phase, show energy fluctuations such as domain walls separating regions of opposite ordering, which are artifacts of non-adiabatic state preparation, and single particle errors due to readout and quantum fluctuations (see the shot in Fig.~\ref{fig:schematic}). 
Nevertheless, the measurement data is capable of providing estimators of correlation functions and susceptibilities accurate enough to extract values for the critical exponents at the WF critical point that are in close agreement with known universal results~\cite{ebadi2021quantum, samajdar2020complex}.
A detailed investigation into how the imperfections in the data manifest in the training of RNN wavefunctions is discussed in Appendix \ref{app:data}. 

%
\section{recurrent neural network wavefunctions}
%

In this work, we employ RNNs to encode ground state wavefunctions~\cite{hibat2020recurrent}.
RNNs are universal function approximators, meaning they can be made expressive enough to encode any arbitrarily complex function by increasing the number of tuneable parameters (or weights, $\mathcal{W}$) in the network~\cite{schafer2006recurrent}. 
Additionally, RNNs belong to a class of generative models known as autoregressive neural networks.
Instead of directly encoding a target joint distribution $p(\mathbf{x})$, these models learn conditional probability distributions over the $N$ individual variables~$p(x_i\vert x_{i-1},x_{i-2}\dots x_{1})$. The chain rule of probabilities allows the overall joint distribution to be decomposed into a sequence of such conditionals, 
\begin{equation}
    p(\mathbf{x}) = p(x_1)p(x_2\vert x_1)\dots p(x_N\vert x_{N-1}\dots x_1).
\label{ProbChainRule}
\end{equation}
This autoregressive property gives rise to a number of desirable features. For instance, the overall normalization of the joint probability distribution is guaranteed by the decomposition into normalized conditionals. 
This normalization means the joint distribution can directly produce independent samples, bypassing the need for Markov chain Monte Carlo sampling. The ability to efficiently obtain independent samples is particularly crucial during variational optimization.

\begin{figure}[htb!]
    \includegraphics[width=\linewidth]{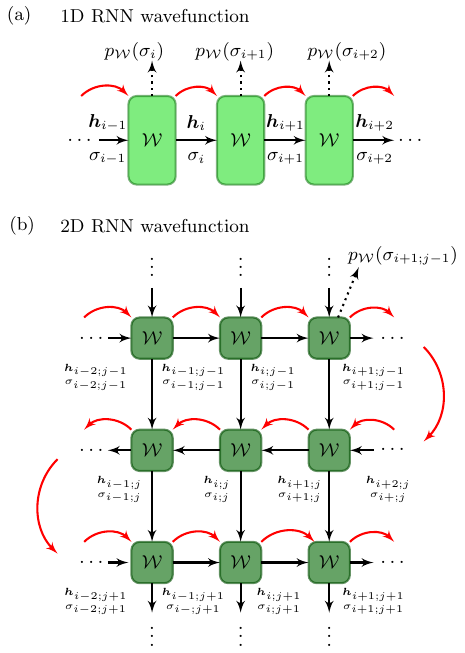}
    \vspace*{-5mm}
    \caption{
    (a) A 1D RNN wavefunction, where each RNN cell (green boxes) takes the hidden vector $\boldsymbol{h}_{i-1}$ and vector containing the state of the previous qubit $\boldsymbol{\sigma}_{i-1}$ and performs a non-linear transformation involving the tuneable parameters $\mathcal{W}$, producing a new hidden vector $\boldsymbol{h}_{i}$ and the probability over the current qubit conditioned on all previously sampled qubits $p_{\mathcal{W}}(\sigma_i\vert\sigma_{<i})$, from which the state of the current qubit $\boldsymbol{\sigma}_i$ can be sampled. The sampling path (red arrows) exactly follows the path in which hidden vectors and state vectors are passed (black arrows).
    (b) A 2D RNN wavefunction, where each RNN cell (green boxes) performs a non-linear transformation on the hidden vectors and state vectors from nearest neighbor RNN cells, which correspond to neighboring qubits. The transformation produces the new hidden vector and conditional probability. Note that while the information is passed in two directions (black arrows), the sampling path is one dimensional (red arrows) as required by the autoregressive property. 
    }
    \label{fig:RNNS}
\end{figure}

The elementary building block of an RNN is a \emph{recurrent cell}, denoted by the green boxes in Fig.~\ref{fig:RNNS}. 
These cells represent a non-linear transformation which involve the tuneable parameters of the model $\mathcal{W}$ and act on the vectors of \emph{hidden units} $\mathbf{h}$, which are the main conduit through which information is passed in the model. In the one-dimensional case, this transformation maps an incoming vector of hidden units $\mathbf{h}_{i-1}$, with dimension $N_h$, and an incoming vector $\boldsymbol\sigma_{i-1}$, which encodes the state of the previous variable, to an output vector of hidden units $\mathbf{h}_{i}$, also with dimension $N_h$.
From the new hidden vector, which has been updated to include information about the newly sampled variable via the non-linear transformation, one can obtain a probability distribution over possible states for the present variable, conditioned on all previously sampled variables $p(\boldsymbol\sigma_i\vert \boldsymbol\sigma_{i<})$. 
This distribution can be sampled for the state of the present variable $\boldsymbol\sigma_i$, which is then passed on to the next recurrent cell. 
This process is naturally represented as a sequence, where $\boldsymbol\sigma_{i-1}$ is sampled first and used to calculate $p(\boldsymbol\sigma_i\vert \boldsymbol\sigma_{i<})$, from which $\boldsymbol\sigma_{i}$ can be sampled.   

The sequence ordering, which is indicated by the red arrows in Fig.~\ref{fig:RNNS}, ensures that the autoregressive property is obeyed.
Fig.~\ref{fig:RNNS} also demonstrates how RNNs can be made multi-dimensional by allowing the recurrent cell to take multiple inputs, which facilitates information passing in more than one direction. It is important, however, that the sampling path (red arrows) remains strictly one-dimensional, so that the autoregressive property is not violated.
This extension of the RNN is especially beneficial for data or systems with multi-dimensional structure, as demonstrated by our results. 
We choose a gated recurrent unit (GRU) as our recurrent cell \cite{cho-etal-2014-learning} for the 1D RNN wavefunction or the higher-dimension Tensorized-GRU~\cite{hibat2022supplementing} for the 2D RNN wavefunction.

In the context of using RNNs as wavefunctions, the target distribution is the probability density function given by the squared amplitudes of the ground state wavefunction, which we refer to as the \emph{Born distribution}. Applying the chain rule of probabilities, the Born distribution can be exactly decomposed into conditional probability distributions over individual qubits, 
\begin{equation}
    \vert\Psi_{GS}(\boldsymbol{\sigma})\vert^2 = p(\boldsymbol{\sigma}) = \prod_i^N p(\sigma_i\vert\sigma_{i-1},\dots,\sigma_1).
\label{ProbChainRuleSpins}
\end{equation}
Each recurrent cell outputs a a hidden vector which is used to compute a conditional probability over a single qubit conditioned on the previously sampled qubits. Therefore, in the case of positive real-valued wavefunctions, such as the ground state of Eq.~\eqref{Hamiltonian}, the distribution encoded in the RNN represents the full quantum state
$\Psi(\boldsymbol\sigma) = \langle \boldsymbol\sigma\vert\Psi\rangle = \sqrt{p_{\text{RNN}}(\boldsymbol\sigma;\mathcal{W})}$.
Further modifications can be used to reconstruct wavefunctions with complex amplitudes~\cite{hibat2020recurrent}. Additionally, modifications can be made such that each recurrent cell corresponds to a ``unit cell'' of qubits, rather than a single qubit, enabling the study of non-Bravais lattices~\cite{hibat2023investigating}.

These models are trained by optimizing the parameters in this network $\mathcal{W}$ such that a specified loss function is minimized. More discussion on the loss function(s) and optimization procedures follow.

%
\section{Data-Enhanced Variational Monte Carlo}
%

We follow the recently-introduced training procedure~\cite{bennewitz2022neural} which we refer to as data-enhanced variational Monte Carlo~\cite{czischek2022data}.
This \emph{hybrid} approach consists of two distinct phases of the training (see Fig.~\ref{fig:schematic}). 
First, our RNN wavefunction is trained in a \emph{data-driven} setting, where we make use of projective measurements of the experimental Rydberg atom array.
The remainder of the training is carried out in a \emph{Hamiltonian-driven} setting, where we employ variational optimization common to machine learning based variational Monte Carlo simulations. 
The only difference between these two settings is the loss function governing the optimization.

During data-driven training, we train our RNN wavefunction in an unsupervised manner to learn the probability distribution over spin configurations that is contained in the data set $\mathcal{D}$. We use the Kullback-Leibler divergence as our loss function, 
\begin{align}
    \mathcal{L}_{\text{DKL}}(\mathcal{W}) &= \sum_{\{\boldsymbol\sigma\}} p_{\mathcal{D}}(\boldsymbol\sigma)\text{log}\,\,\frac{p_{\mathcal{D}}(\boldsymbol\sigma)}{p_{\text{RNN}}(\boldsymbol\sigma;\mathcal{W})} \label{KLDiv1}\\
    &\approx - S_{\mathcal{D}} - \frac{1}{\vert\mathcal{D}\vert} \sum_{\boldsymbol\sigma \in \mathcal{D}} \text{log}\,\, p_{\text{RNN}}(\boldsymbol\sigma;\mathcal{W}),
    \label{KLDiv2}
\end{align}
where we have introduced the entropy of the data set $S_{\mathcal{D}}=-\sum_{\{\boldsymbol\sigma\}}p_{\mathcal{D}}(\boldsymbol\sigma)\text{log}\,p_{\mathcal{D}}(\boldsymbol\sigma)$ and approximated $p_{\mathcal{D}}$ with the empirical distribution, which reduces the sum over all configurations $\{\boldsymbol{\sigma}\}$ to the sum over the data in $\mathcal{D}$. 
It is clear from Eq.~\eqref{KLDiv1} that the minimum of $\mathcal{L}_{\text{DKL}}$ occurs when the two probability distributions, $p_{\mathcal{D}}(\boldsymbol\sigma)$ and $p_{\text{RNN}}(\boldsymbol\sigma;\mathcal{W})$, are equal. This is achieved in practice by determining the parameters $\mathcal{W}$ of our RNN wavefunction that minimize this loss, which we approximate with Eq.~\eqref{KLDiv2}.
With the availability of large (informationally-complete) amounts of data, the empirical distribution is a good approximation to the Born distribution and the KL loss could be employed on its own to reliably reconstruct a quantum state of interest~\cite{torlai2018neural}. 
We focus on the scenario where one has access to only a limited amount of data and full tomography or even data-driven quantum state reconstruction is out of reach. 
Thus, we treat the data-driven training as a type of \emph{pre-training} which improves the performance of the subsequent variational optimization.

In the Hamiltonian-driven setting, the same RNN wavefunction can be trained to represent the ground state of the quantum system without making use of data, and instead using the Hamiltonian that is presumed to describe the system. Here, we use standard variational optimization in which the expectation value of the energy is minimized~\cite{becca_sorella_2017}. This optimization technique can be used to target ground states because of the variational principle, which guarantees that the energy will not converge to a value below the true ground state of the Hamiltonian. We define the new loss function for variational optimization as
\begin{align}
    \mathcal{L}_H(\mathcal{W}) &= H_{\text{RNN}} \\
    &= \frac{\langle \Psi_{\mathcal{W}}\vert \hat{H}\vert \Psi_{\mathcal{W}}\rangle}{\langle \Psi_{\mathcal{W}}\vert \Psi_{\mathcal{W}}\rangle}\\
    &= \frac{1}{N_s}\sum_{\boldsymbol\sigma \sim p_{\text{RNN}}(\boldsymbol\sigma;\mathcal{W})} H_{\text{loc}}(\boldsymbol\sigma),
\end{align}
where we introduce the local energy,
\begin{equation}
    H_{\text{loc}}(\boldsymbol\sigma) = \frac{\langle \boldsymbol\sigma\vert \hat{H}\vert \Psi_{\mathcal{W}}\rangle}{\langle \boldsymbol\sigma\vert \Psi_{\mathcal{W}}\rangle}.
\end{equation}
Here, we approximate the ground state energy of our RNN wavefunction $H_{\text{RNN}}$ by averaging the local energy $H_{\text{loc}}$ evaluated on $N_s$ samples drawn from $p_{\text{RNN}}(\boldsymbol\sigma;\mathcal{W})$. This quantity is treated as our loss function and is minimized during Hamiltonian-driven training. A more detailed discussion of this approximation of $H_{\text{RNN}}$ and the local energy can be found in Appendix \ref{app:vmc}.

%
\section{Results}
%

In this section, we demonstrate the successes of data-enhanced Variational Monte Carlo as a method for training both one-dimensional (1D) and two-dimensional (2D) RNN wavefunctions. In what follows, we benchmark our results against data provided by quantum Monte Carlo (QMC) simulations~\cite{merali2021stochastic}. For Rydberg atom arrays described by the  Hamiltonian Eq.~\eqref{Hamiltonian}, QMC is free of the sign problem, and can therefore provide groundstate observables that are exact to within statistical errors. Additionally, QMC can provide unbiased (but autocorrelated) samples of groundstate configurations in the Rydberg occupation basis, which we use to train RNN generative models in Appendix~\ref{app:data}.

In our main results, we consider the experimental data for the disordered-to-checkerboard phase transition from Ref.~[\onlinecite{ebadi2021quantum}] for a $16 \times 16$ array of $^{87}$Rb atoms. 
The experimental parameter governing this phase transition is the detuning $\delta$; we report our results in terms of the dimensionless ratio $\delta/\Omega$. 
We consider 31 values of $\delta/\Omega$ ranging from roughly $-0.4$ to $3.2$ with the phase transition occurring at $\delta_c/\Omega = 1.12(4)$~\cite{ebadi2021quantum}.
In all of the following, we fix $N_h = 2L$ for the 1D RNN, which was previously shown to equip these models with sufficient expressive capabilities~\cite{czischek2022data}. For the 2D RNN, we set $N_h = L$, keeping the total number of parameters in these two types of ans\"atz roughly equal. 
We initialize our RNN wavefunction using the methods introduced by Glorot \emph{et al.}~\cite{pmlr-v9-glorot10a}. Then, using the Adam optimizer~\cite{kingma2014adam}, which is an extension to stochastic gradient descent with an adaptive learning rate, we optimize the tuneable parameters $\mathcal{W}$ according to the specified loss function.
Additional information about the experimental parameters and hyperparameters of our RNN wavefunctions can be found in Appendix \ref{app:params}.

As discussed above and illustrated in Fig.~\ref{fig:schematic}, training proceeds through two different phases; data-driven then Hamiltonian-driven,
the difference being the loss function which is minimized. This hybrid training scheme involves simply switching between loss functions after a specified number of data-driven training steps given by $t_{\text{trans}}$.
Here, we fixed $t_{\text{trans}} = 1000$ for our hybrid trained 1D RNN wavefunctions and $t_{\text{trans}}=100$ for our hybrid trained 2D RNN wavefunctions based on observations that after this many data-driven training steps, the respective models had learned from the measurement data without overfitting to it. 
Note, in this work we only explore one phase of data-driven training, followed by one phase of Hamiltonian-driven training.  Other transition schedules are possible; however we observe this order of training to give the best results, consistent with past studies using perfect synthetic data~\cite{czischek2022data}.

\subsection{Improving variational Monte Carlo simulations with data-enhancement} 

\begin{figure*}[!]
    \centering
    \includegraphics{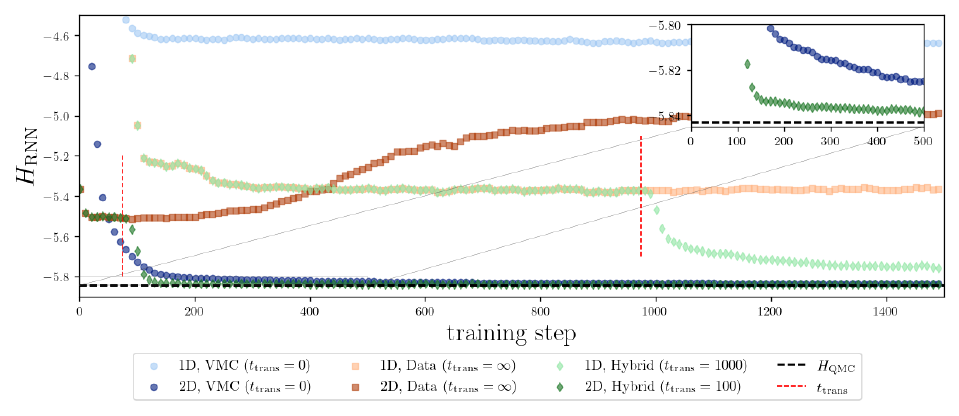}
    \vspace*{-5mm}
    \caption{
    Convergence of the estimated ground state energy $H_{\mathrm{RNN}}$ for 1D and 2D RNN wavefunctions trained in a purely Hamiltonian-driven setting (blue), a purely data-driven setting (orange), and in the hybrid data- then Hamiltonian-driven setting (green). The target state is deep in the checkerboard phase ($\delta/\Omega = 3.173$), meaning it is highly ordered. The ground state energy given by quantum Monte Carlo is marked by the black dashed line. For hybrid training, the transition from data-driven to Hamiltonian-driven training is marked by the red dashed line. 
    }
    \label{fig:trainingcurves}
\end{figure*}

\begin{figure*}[!]
    \centering
    \includegraphics{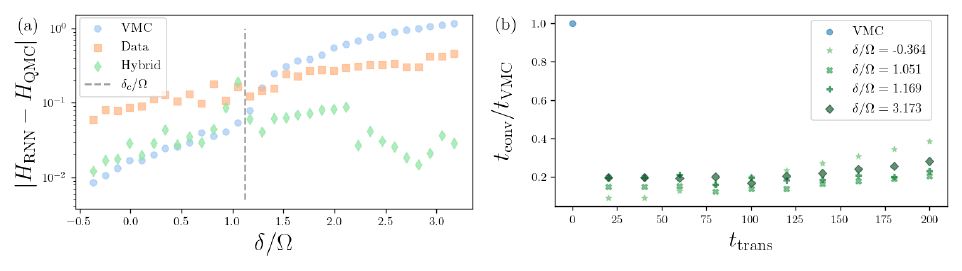}
    \vspace*{-6mm}
    \caption{
    (a) The absolute error between the best estimated ground state energy achieved by trained 1D RNN wavefunctions $H_{\mathrm{RNN}}$ and the ground state energy given by quantum Monte Carlo simulations $H_{\mathrm{QMC}}$. 
    (b) Convergence time $t_{\mathrm{conv}}$ as a function of the transition point $t_{\mathrm{trans}}$ in the hybrid training of 2D RNN wavefunction. The convergence time has been rescaled by the time it takes for the RNN to be trained purely variationally $t_{\mathrm{VMC}}$, showing the relative speedup provided by data-enhancement. This is because $t_{\mathrm{conv}}$ and $t_{\mathrm{VMC}}$ depend on the value of $\delta/\Omega$.
    }
    \label{fig:generalresults}
\end{figure*}

To illustrate the behavior of our RNN wavefunctions during the training process, Fig.~\ref{fig:trainingcurves} shows the training curves for 1D and 2D RNN wavefunctions optimized to capture the ground state of our Rydberg system deep in the checkerboard phase ($\delta/\Omega = 3.173$).
We show the results of data-enhanced VMC with a transition time $t_{\text{trans}}$, and compare it to a purely Hamiltonian-driven approach ($t_{\text{trans}} = 0$), as well as a purely data-driven approach ($t_{\text{trans}} = \infty)$.
These training curves epitomize the success of data-enhanced VMC for both the 1D and 2D RNN wavefunctions. 
For the 1D RNN, variational optimization struggles and the model is unable to capture the properties of the target ground state. The use of the experimental data to pre-train the model significantly improves the overall performance, as seen through the converging ground state energy estimate $H_{\text{RNN}}$. For the 2D RNN, the models trained purely variationally and with data-enhanced VMC reach comparable and accurate ground state energies, but the data-enhancement provides a speed up in the overall time to convergence, as seen in the inset of Fig.~\ref{fig:trainingcurves}. 
We also point out the significant gap between the ground state energy obtained from our QMC simulations and the energies achieved by both the 1D and 2D RNN wavefunctions trained with data alone. In Appendix~\ref{app:data} we show that this gap is at least partly due to both the limited and the imperfect nature of the experimental data. Fig.~\ref{fig:trainingcurves} illustrates these training dynamics for the case where we are learning a ground state in the ordered checkerboard phase, but we observe that these trends also hold more broadly across the phase transition.
Fig.~\ref{fig:generalresults}a and Fig.~\ref{fig:generalresults}b illustrate the outcomes of data-enhanced VMC for both the 1D RNNs and 2D RNNs
for a variety of detuning values on either side of the critical coupling $\delta_c/ \Omega$.

Fig.~\ref{fig:generalresults}a shows that, in some cases, the energies achieved by the 1D RNN are improved by over an order of magnitude when data-enhanced VMC is used as the training procedure, particularly 
in the ordered checkerboard regime.
A similar effect was not observed in Ref.~[\onlinecite{czischek2022data}] because the method was tested at only one point in the phase diagram, close to a phase transition. The results here are consistent, in that the performance of variational optimization and data-enhanced VMC are comparable in the disordered phase and around the phase transition, with data-enhancement still providing a speedup in convergence time in these regimes. 
In addition to the accuracy improvements seen in the ground state energy estimates in the ordered regime (Fig.~\ref{fig:generalresults}a), we find that the use of data allows the 1D RNN wavefunction to capture the 2D ordering of the checkerboard phase, which it is unable to do when trained using the Hamiltonian alone. 
This might be understood since the data set, which is a set of projective measurement outcomes that are exemplary of the ground state or states close to it, contains a more explicit representation of the correlations between atoms in the array. 
Using this data in the earliest epochs of training optimizes the model to directly reflect this information. This early-phase training seems to be crucial in moving the gradient descent algorithm into a smaller subspace that can more easily be optimized by later-phase variational training. 
More details and discussion can be found in Appendix~\ref{app:stagmagtraining}.

Fig.~\ref{fig:generalresults}b shows that for the 2D RNN wavefunctions, the data-enhancement consistently provides a significant speedup in time-to-convergence for varying points across the phase transition. 
For any $t_{\text{trans}} > 0$ the time to convergence is reduced, with the most significant speed up occurring when $t_{\text{trans}}$ is equal to the point at which the minimum energy is achieved during data-driven training. While these results show that $t_{\text{trans}}$ could be optimized for each trained model, they also motivate our choice for a uniform transition time, since appreciable speedups are observed for all transition times.

It should be noted that $t_{\text{conv}}$ is rescaled by $t_{\text{VMC}}$, which masks the absolute speedup in convergence time. As an example, at $t_{\text{trans}}=80$ the values of $t_{\text{conv}}/t_{\text{VMC}} = 87/541,186/1498,267/1667,179/883$ for the four values of $\delta/\Omega$ in increasing order. Both $t_{\text{conv}}$ and  $t_{\text{VMC}}$ vary for the different points across the phase transition, with the longest convergence times occurring for the values of $\delta/\Omega$ close to the phase transition. While the relative speedups are comparable, the absolute speedups show that the data offers the largest gain in the vicinity of the phase transition, where it is expected in the thermodynamic limit that the correlation length will diverge and the energy gap will vanish.

%
\subsection{Calculating observables from trained RNN wavefunctions}
%

In the previous section, we focused on demonstrating how data-enhanced VMC is a successful training protocol that enables both the 1D and 2D RNN wavefunctions to more efficiently learn quantum ground states.
Here, we aim to further demonstrate that trained RNN wavefunctions are accurate representations of ground states of the Rydberg array across the quantum phase transition of interest. We do so by using our trained RNN wavefunctions to calculate physically relevant ground-state observables across the transition. 

We define first a simple order parameter for the checkerboard phase, 
\begin{equation}
    M_s = \frac{1}{2}\sum_{i\in A}\hat{\sigma}^z_i - \frac{1}{2}\sum_{j \in B}\hat{\sigma}^z_j,
\label{stagmag}
\end{equation}
where $A$ and $B$ refer to the two sublattices of the bipartite square lattice, and we define
the Pauli operator $\hat{\sigma}^z_i = 2 \hat{n}_i-1$.
This is equivalent to the {\it staggered magnetization} in a spin-1/2 system, and we use that nomenclature henceforth.
In the phase described by this order parameter, the Fourier transform of experimental measurement outcomes of the Rydberg occupation
would result in a sharp peak at wavevector $\bold{k} = (\pi,\pi)$~\cite{ebadi2021quantum}.

Because Eq.~\eqref{stagmag} is diagonal in the occupation basis, this quantity can be easily calculated directly from samples or projective measurements. For example, the staggered magnetization can be calculated from samples drawn from a trained RNN wavefunction. The ensemble average then defines the estimator,
\begin{equation}
    \langle M_s \rangle_{\text{RNN}} = \frac{1}{N_s} \sum_{\boldsymbol\sigma \sim p_{\text{RNN}}(\boldsymbol\sigma;\mathcal{W})} \vert M_s(\boldsymbol\sigma)\vert ,
\label{expectation_stagmag}
\end{equation}
which can provide an important quantity to verify the RNN wavefunction accuracy past the standard loss function or the energy.

\begin{figure}[!]
    \centering
    \includegraphics[width=\linewidth]{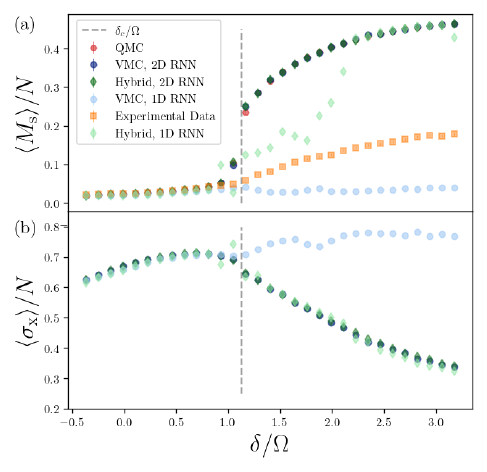}
    \vspace*{-8mm}
    \caption{
    (a) The order parameter for the checkerboard phase estimated from quantum Monte Carlo simulations, the trained RNN wavefunctions, and the raw experimental measurement data.
    (b) The spatially averaged expectation value of $\sigma^x$ estimated from the trained RNN wavefunctions. The standard error is shown, but the associated error bars are small and hidden by the data markers.
    }
    \label{fig:observables}
\end{figure}

While the estimation of $\langle M_s \rangle_{\text{RNN}}$ does allow us to verify how well our models have captured the phase across the phase transition, the staggered magnetization can also be calculated directly from experimental measurement data or easily tracked in most world-line QMC simulations.
On the contrary, estimators for off-diagonal operators are harder to obtain experimentally.
Because our trained RNN wavefunctions provide explicit access to the learned approximation of the ground state wavefunction, we have a representation of the associated Born distribution that can then be probed easily. This allows for an efficient estimation of off-diagonal observables. 

In order to demonstrate how the trained RNN wavefunctions can be used to estimate off-diagonal observables, we consider the spatially averaged expectation value of $\hat\sigma^x$, which can be defined as
\begin{equation}
    \langle\hat{\sigma}^x\rangle_{\text{RNN}} = \frac{1}{N_s}\sum_{\boldsymbol\sigma\sim p_{\text{RNN}}(\boldsymbol\sigma;\mathcal{W})}  \frac{1}{N}\sum_{i=1}^{N} \frac{ \Psi_{\mathcal{W}}(\boldsymbol\sigma_i\prime)}{\Psi_{\mathcal{W}}(\boldsymbol\sigma)},
\label{expectation_sigmax}
\end{equation}
where $ (\boldsymbol\sigma_i\prime)_j = \boldsymbol\sigma_j - \delta_{i,j}$, meaning $\boldsymbol\sigma_i\prime$ is equivalent to $\boldsymbol\sigma$ except that the {\it i}-th spin is flipped. The inner sum is the spatial average and the outer sum is the ensemble average used to estimate the expectation value. Having an explicit representation of our approximation of the Born distribution is essential for calculating the ground state wavefunction amplitude for specific configurations (in this case, e.g. $\boldsymbol\sigma_i\prime$). For more details on estimating off diagonal operators from trained RNN wavefunctions, see Appendix \ref{app:offdiag}.

Fig.~\ref{fig:observables} displays these two observables, estimated for all values of $\delta/\Omega$ for which we have experimental data.
The upper panel shows the staggered magnetization as a function of $\delta/\Omega$. As expected, the 2D RNN wavefunctions, trained purely variationally and with data-enhanced VMC, are in close agreement with the staggered magnetization estimators obtained through QMC, which we take to be the ground truth. 
On the other hand, 
the hybrid trained 1D RNN wavefunctions show much better performance than those trained with standard VMC. This performance is consistent with the behavior seen in Fig.~\ref{fig:generalresults}a in that there is more error in the vicinity of the phase transition, but better agreement with QMC in the highly ordered phase. Estimates of $M_s$ obtained from the raw experimental data set are also shown as a means of demonstrating that while neither the experimental data nor the 1D RNN trained with standard VMC are able to accurately capture the staggered magnetization across the phase transition, combining the two leads to significantly higher accuracy.

The lower panel shows the spatially averaged expectation value of $\hat\sigma_x$. 
These results again demonstrate how the 1D RNN wavefunctions trained with only VMC are unable to capture the transition to the ordered phase. Interestingly, the hybrid trained 1D RNN is in close agreement with the 2D RNN, even in the region around the phase transition.
These results show how even sub-optimal models (in this case the 1D RNN attempting to capture 2D ordering patterns) can be drastically improved by pre-training with raw experimental data.

%
\section{Conclusions and Outlook}
%

In this paper we use raw experimental data from a 256-qubit 
Rydberg atom array, to pre-train state-of-the-art autoregressive recurrent neural network (RNN) wavefunction ans\"atze. 
The data contains shots of projective measurements taken in the Rydberg occupation basis across the disordered-to-checkerboard phase transition, obtained in Ref.~[\onlinecite{ebadi2021quantum}].
We demonstrate that subsequent variational training of the wavefunction, guided by Eq.~\eqref{Hamiltonian}, evolves it towards the ground state of the presumptive Rydberg Hamiltonian despite the limited size and imperfections of the data set used for pre-training.

RNNs -- first developed for applications in natural language processing -- are traditionally one dimensional (1D), and have been demonstrated to successfully support data-enhanced variational Monte Carlo (VMC) in the past~\cite{hibat2020recurrent,czischek2022data}.  For our two-dimensional (2D) $16 \times 16$ Rydberg array, we find that 1D RNNs make poor variational wavefunctions when trained with a Hamiltonian-driven strategy alone, with deteriorating performance as the 2D checkerboard order of the system becomes more pronounced. However, pre-training with a data-driven strategy, using raw experimental shots, results in a significantly improved variational ground state.
We quantify this by estimating the energy and other relevant observables across the disordered-to-checkerboard transition.
In particular, we calculate both the checkerboard order parameter (the qubit staggered $\sigma^z$ magnetization) and the spatially averaged expectation value of $\sigma^x$, which is inaccessible from the raw experimental data in the occupation basis. These quantities all reflect vast qualitative improvement, particularly in the checkerboard phase, when obtained with our data- then Hamiltonian-driven hybrid optimization.

We further find that the use of the experimental data quite generally provides a speed up in the convergence time of subsequent variational optimization in a variety of RNN wavefunctions. For the 1D RNN this includes the aforementioned qualitative improvement in observables, suggesting the experimental data helps drive the RNN parameters towards a fundamentally different optimum during training. 
In this context, the pre-training of our ans\"atze with experimental data can be viewed as a means for better initializing the parameters of the wavefunction prior to the subsequent variational optimization. From this perspective, our results are consistent with the previous body work in the machine learning literature that shows that the initialization of neural networks has tremendous effects on convergence~\cite{pmlr-v9-glorot10a,pretrain}, which presently motivates significant research efforts in determining optimal methods for initialization~\cite{initializationreview}.

We also explored state-of-the-art 2D RNN wavefunctions which are more suited to the 2D Rydberg array geometry, and found that their Hamiltonian-driven optimization was highly effective, especially when compared to the variationally trained 1D RNN wavefunctions. Despite this, we are able to demonstrate that pre-training with experimental data also provides clear value, in that the time-to-convergence of the variational optimization is significantly improved.
Specifically, pre-training on just 1000 experimental shots reduced the time-to-convergence by up to $\mathcal{O}(10^3)$ training iterations -- an improvement that can translate into a significant reduction of computer time for an RNN model of this size.
While 2D RNN wavefunctions converge in fewer training steps than their 1D counterparts, Rydberg atom arrays continue to scale towards thousands of atoms and the size of the models used to characterize them must be scaled in response.  
In this limit, even 2D RNNs may require long convergence times using Hamiltonian-driven training alone. 
Here, pre-training with experimental data (even limited, imperfect data) could appreciably improve the variational ground state, and lessen the task of designing clever ans\"atze for the large Rydberg arrays of the near future.

Furthermore, since these models can be trained efficiently to represent quantum ground states in various phases with excellent accuracy, we suggest that they could be useful tools for experimentalists in need of quick characterization as systems are scaled in size or used to explore new regimes of quantum matter. While data-enhanced VMC simulations cannot replace the tomographic reconstruction of quantum states, this method provides an alternative that is rooted in the experiment without requiring prohibitive amounts of data that will grow with increasing system size. 

Most importantly, our results pave the way for hybrid quantum-classical simulation of more complex systems, where variational optimization becomes challenging and quantum data might be essential for capturing the target quantum state. 
Indeed, RNNs have been demonstrated as powerful wavefunction ans\"atze for learning ground states with more exotic structures such as non-trivial signs~\cite{hibat2022supplementing} and topological order~\cite{hibat2023investigating}.
It is conceivable that there are many instances where our hybrid strategy will be essential for RNNs and other autoregressive models to accurately capture these and other complex correlations in  quantum matter. 
Particularly promising examples include systems with sign or phase structures, such as frustrated Hamiltonians as could be realized with XXZ spin interactions \cite{XXZ}, and systems undergoing dynamical evolution. In these cases, measurements in multiple bases used for early-epoch training could significantly help subsequent variational optimization learn the wavefunction's complex phase, a task that is believed to be more difficult than learning amplitudes alone~\cite{westerhout2019neural,bukov2021learning}. 
In addition, in experiments where the state preparation protocol might lead to metastable states, a subsequent variational optimization could be used to give information about the distance to the target groundstate. Ultimately, this hybrid procedure for error mitigation could be developed into a more formal error correcting protocol \cite{TorlaiDecoder,Chamberland_2018,bennewitz2022neural}.  These results hint at the promise of data-enhanced variational Monte Carlo simulations for leveraging data from today's quantum devices and contributing to their path towards quantum advantage in the future.

%
\section*{Acknowledgements}
%
We thank J. Carrasquilla, R. Wiersema, and S. Czischek for their insights and discussion, and in particular E. Merali for his invaluable support with the quantum Monte Carlo simulations.
We would also like to thank H. Levine, A. Keesling, G. Semeghini, A. Omran, D. Bluvstein for the use of experimental data presented in this work.
This research was enabled in part by computational support provided by the Shared Hierarchical Academic Research Computing Network (SHARCNET) and the Digital Research Alliance of Canada.

We acknowledge financial support from the 
US Department of Energy (DOE Quantum Systems Accelerator Center, contract number 7568717 and DE-SC0021013),  the Center for Ultracold Atoms, the National Science Foundation, the Army Research Office MURI (grant number W911NF-20-1-0082), the DARPA ONISQ program (grant number W911NF2010021),
the Natural Sciences and Engineering Research Council of Canada (NSERC),
the Deutsche Forschungsgemeinschaft (DFG, German Research Foundation) under Germany’s Excellence Strategy—EXC-2111—390 814868, and the Perimeter Institute.
Research at Perimeter Institute is supported in part by the Government of Canada through the Department of Innovation, Science and Economic Development Canada and by the Province of Ontario through the Ministry of Economic Development, Job Creation and Trade.

\appendix
\section{Experimental Rydberg Atom Array Data}\label{app:data}
In this section we aim to answer three main questions concerning the experimental data set we consider and its use in training our RNN wavefunctions. First, how do the imperfections in the experimental data set manifest in the training of our RNN wavefunctions? Second, how does having a limited amount of measurement data hinder the training of our RNN wavefunction? And finally, does this fully explain why our RNN wavefunctions are unable to accurately capture the ground states with data alone?

\begin{figure}[b]
    \centering
    \includegraphics{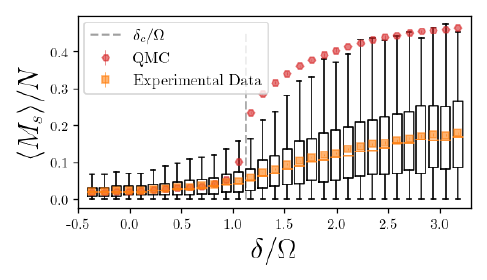}
    \vspace*{-5mm}
    \caption{
    The checkerboard order parameter estimated from quantum Monte Carlo simulations (red) and the raw experimental measurement data. The black box plots show the full distribution of order parameters obtained from individual shots of projective measurements. Half of the values are contained within the boxes, which are defined by the upper and lower quartiles of the distribution. The bars outside of these boxes show the range of the remaining values with outliers excluded. 
    While some of the shots produce values of the order parameter that are in agreement with our quantum Monte Carlo simulations, the median (orange line) and mean (orange boxes), are lower than the average QMC values after the phase transition where the variance among the data also becomes much higher.  
    }
    \label{fig:whisker}
\end{figure}

The main ``imperfections'' in the projective measurement data we consider are the fluctuations which are present because the experimentally prepared states are \emph{low-energy states}. In the experiments, perfect ground states are not prepared due coherent perturbations, such as parameter mis-calibration and non-adiabatic effects due to finite preparation time, as well as due to incoherent errors. 
While these fluctuations could be mitigated by careful experimental fine-tuning and potentially data post-processing, we aim to show that even with such imperfections, the experimental data is valuable when used in our hybrid training approach.
Deep in the ordered phase, the energy fluctuations in the considered data often appear as domain walls that separate regions of the array with opposite anti-ferromagnetic ordering (see the experimental shot shown in Fig.~\ref{fig:schematic}). Having regions of opposite ordering is destructive in the calculation of the staggered magnetization, Eq.~\eqref{stagmag}. This explains why, in Fig.~\ref{fig:observables}, the average values of the staggered magnetization estimated directly from the experimental data are not in close agreement with the values obtained from our simulations. 
Figure~\ref{fig:whisker} instead compares the entire distribution of staggered magnetization values obtained from the experimental data to the values obtained from our QMC simulations. From this, it is clear that some of the projective measurements captured the exact many-body ground state, or a state very close to it. 

\begin{figure}[]
    \centering
    \includegraphics{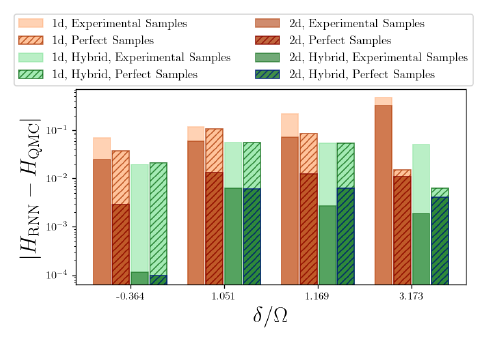}
    \vspace*{-5mm}
    \caption{
    The absolute error between the best estimated ground state energy achieved by trained RNN wavefunctions $H_{\mathrm{RNN}}$ and the ground state energy given by quantum Monte Carlo simulations $H_{\mathrm{QMC}}$ for four values of $\delta/\Omega$ that span the phase transition. We compare models trained with only the experimental data to models trained with perfect samples obtained from quantum Monte Carlo simulations (orange and striped orange, respectively). We also compare models trained with the hybrid data- then Hamiltonian-driven method when the experimental data and the perfect samples are used for data-driven pre-training (green and striped green, respectively). Our experimental data contains 1,000 snapshots of projective measurements and we equivalently use 1,000 uncorrelated QMC samples as our ``perfect'' samples.
    }
    \label{fig:expvsqmc}
\end{figure}

Having established that there are imperfections present in the experimental data set, it is not yet clear how those imperfections manifest in the training of a model. In order to explore their effects, we compare models trained on 1,000 experimental samples to models trained on 1,000 perfect, uncorrelated samples obtained from our QMC simulations. In Fig.~\ref{fig:expvsqmc}, we compare both 1D and 2D RNN wavefunctions trained variationally and using our hybrid approach at four points across the phase transition.

The first thing to observe in Fig.~\ref{fig:expvsqmc} is that the imperfections in the experimental data have a clear effect on the training outcome when only data-driven training is employed. This is evidenced by the fact that the striped orange bars are lower than the plain orange bars for all values of delta and both types of RNN wavefunctions. When the hybrid approach is used to train our models, however, the energy estimates are more accurate in all cases, as seen by comparing the green bars to the corresponding orange bars. Furthermore, after the subsequent variational optimization, the effects of the imperfections in the data are less evident. When the hybrid approach was employed, the models pre-trained with the experimental data performed as well as those pre-trained on the perfect QMC samples. These results can be observed by comparing the striped green bars to the plain green bars. This is related to the idea that VMC can be seen as means for smoothing out errors and noise in data obtained from quantum simulations~\cite{bennewitz2022neural}. Importantly, this result emphasizes the value of the data, despite the imperfections, and suggests that data with even more non-adiabatic energy excitations (i.e. data from experiments using faster sweep rates) could still provide value when used for data-enhanced VMC simulations.

\begin{figure}[]
    \centering
    \includegraphics{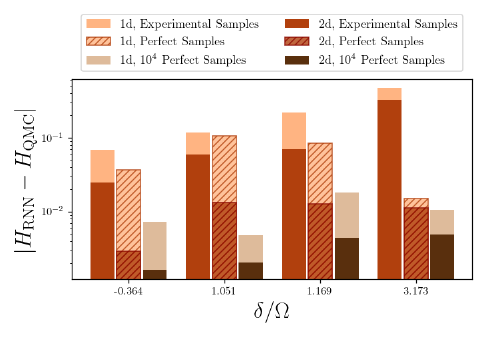}
    \vspace*{-5mm}
    \caption{
    The absolute error between the best estimated ground state energy achieved by trained RNN wavefunctions $H_{\mathrm{RNN}}$ and the ground state energy given by quantum Monte Carlo simulations $H_{\mathrm{QMC}}$ for four values of $\delta/\Omega$ that span the phase transition. We compare models trained with only the experimental data to models trained with $10^3$ perfect samples obtained from quantum Monte Carlo simulations (orange and striped orange, respectively). We compare these models to models trained with $10^4$ perfect samples (brown) to show that additional data can improve the energy error after data-driven pre-training. This corresponds to a better initialization for the subsequent variational optimization. 
    }
    \label{fig:1kvs10k}
\end{figure}

To understand how a \emph{limited} data set impacts the training of an RNN wavefunction, we compare models trained on 1,000 perfect samples obtained from our QMC simulations to models trained on 10,000 perfect samples. We also show the results of the models trained on 1,000 experimental samples for reference. In Fig.~\ref{fig:1kvs10k}, we again compare both 1D and 2D RNN wavefunctions for four values of delta across the phase transition, this time only examining the purely data-driven setting.

Clearly, access to a larger data set systematically improves the accuracy of the trained models for both the 1D and 2D RNN wavefunctions and across the phase transition. This is reflected by the comparison of the brown bars to the orange bars. In the case of the 2D RNN wavefunction, the 10-fold increase in the number of samples results in an accuracy improvement of an order of magnitude (or more). Therefore, the fact that our experimental data set is limited in size does play a role in the data-driven training of our models. This is understandable, as these measurement shots are Born distributed and having more shots means that the empirical distribution in the data set is a better approximation of the true underlying Born distribution. 

From these experiments, it is evident that the imperfect and limited nature of the considered experimental data set does surface during the training (or pre-training) of our RNN wavefunctions. This partially explains why, in Fig.~\ref{fig:trainingcurves}, neither the 1D nor the 2D RNN wavefunctions trained on data alone are able to reach energies close to the energy obtained from our QMC simulation. Our hybrid training method, however, is not hindered by the limited or imperfect nature of the data because after the pre-training, the presumed Hamiltonian is introduced and optimization is continued variationally. This underscores the idea that unprocessed quantum data from today's devices can provide immense value through this type of hybrid quantum-classical simulation.

There is a final possible explanation for the gap in accuracy between energies obtained from RNNs trained on experimental data alone and the energies obtained from our QMC simulations: the assumption of the incorrect Hamiltonian governing the experiment. It is conceivable that either the form of the Hamiltonian is incorrect or the values of the parameters in the Hamiltonian are incorrect. In this case, variational training with a presumed Hamiltonian will guide the RNN wavefunction to the ground state of that Hamiltonian, potentially imposing an incorrect bias. Unfortunately, exploring this possibility is beyond the scope of this work, as it would require much more data than we have access to. We leave such explorations for future work.

\section{Deriving $H_\text{loc}$ for Variational Optimization}\label{app:vmc}
During the variational optimization of our RNN wavefunctions, the tuneable parameters of the model $\mathcal{W}$ are optimized by minimizing the expectation value of the energy, which is given by
\begin{equation*}
    H_{\text{RNN}} = \frac{\langle \Psi_{\mathcal{W}}\vert \hat{H}\vert \Psi_{\mathcal{W}}\rangle}{\langle \Psi_{\mathcal{W}}\vert \Psi_{\mathcal{W}}\rangle}.
\end{equation*}
Recall that our RNN wavefunctions encode a probability distribution $p_{\text{RNN}}(\boldsymbol\sigma;\mathcal{W})$ that yields an approximation to the ground state wavefunction 
$\Psi_{\mathcal{W}}(\boldsymbol\sigma) = \langle \boldsymbol\sigma\vert\Psi_{\mathcal{W}}\rangle = \sqrt{p_{\text{RNN}}(\boldsymbol\sigma;\mathcal{W})}$.
Inserting the identity in the numerator and the denominator, and multiplying by 1, the above expectation value can be equivalently written as
\begin{equation*}
    H_{\text{RNN}} = \frac{\sum_{\boldsymbol\sigma} \vert\langle \Psi_{\mathcal{W}}\vert\boldsymbol\sigma\rangle\vert^2 \times \frac{\langle \boldsymbol\sigma\vert \hat{H}\vert \Psi_{\mathcal{W}}\rangle}{\langle \boldsymbol\sigma\vert \Psi_{\mathcal{W}}\rangle}}{\sum_{\boldsymbol\sigma\prime} \vert\langle \Psi_{\mathcal{W}}\vert\boldsymbol\sigma\prime \rangle \vert^2 }.
\end{equation*}
One can recognize 
$\frac{\vert\langle \Psi_{\mathcal{W}}\vert\boldsymbol\sigma\rangle\vert^2}{\sum_{\boldsymbol\sigma\prime} \vert\langle \Psi_{\mathcal{W}}\vert\boldsymbol\sigma\prime \rangle \vert^2}$
as a normalized probability $P(\boldsymbol\sigma)$. 
Here we also define the local energy,
\begin{equation*}
    H_{\text{loc}}(\boldsymbol\sigma) = \frac{\langle \boldsymbol\sigma\vert \hat{H}\vert \Psi_{\mathcal{W}}\rangle}{\langle \boldsymbol\sigma\vert \Psi_{\mathcal{W}}\rangle}.
\end{equation*}
Approximating $P(\boldsymbol{\sigma})$ with the frequency with which a given configuration $\boldsymbol\sigma$ is sampled from the probability distribution encoded in our RNN wavefunction $p_{\text{RNN}}(\boldsymbol\sigma;\mathcal{W})$, we can define the expectation value of the energy as
\begin{equation*}
    H_{\text{RNN}} \approx \frac{1}{N_s}\sum_{\boldsymbol\sigma\sim p_{\text{RNN}}(\boldsymbol\sigma;\mathcal{W})} H_{\text{loc}}(\boldsymbol\sigma).
\end{equation*}
We minimize this quantity during Hamiltonian-driven training.
Because this expectation value is estimated at each Hamiltonian-driven training step and relies on samples drawn from our model, it is essential that these samples can be taken efficiently. Fortunately, due to the autoregressive nature of RNN wavefunctions, any number of independent samples can be drawn directly from the learned conditional probabilities in one pass of the RNN, which is an advantage over models that require Markov Chain Monte Carlo sampling. 

\section{Expectation Values of Off-Diagonal Operators}\label{app:offdiag}
In this work, we estimate the expectation values of an off-diagonal operator $\hat{\mathcal{O}}$ using our trained RNN wavefunctions. As done for the estimation of the expectation value of the energy, we start with the following form of the expectation value,
\begin{equation*}
    \langle\hat{\mathcal{O}}\rangle_{\text{RNN}} = \frac{\langle \Psi_{\mathcal{W}}\vert \hat{\mathcal{O}}\vert \Psi_{\mathcal{W}}\rangle}{\langle \Psi_{\mathcal{W}}\vert \Psi_{\mathcal{W}}\rangle}.
\end{equation*}
Following the same steps outlined in the Appendix B, this expectation value can be equivalently written as 

\begin{figure*}[ht]
    \includegraphics[width=\linewidth]{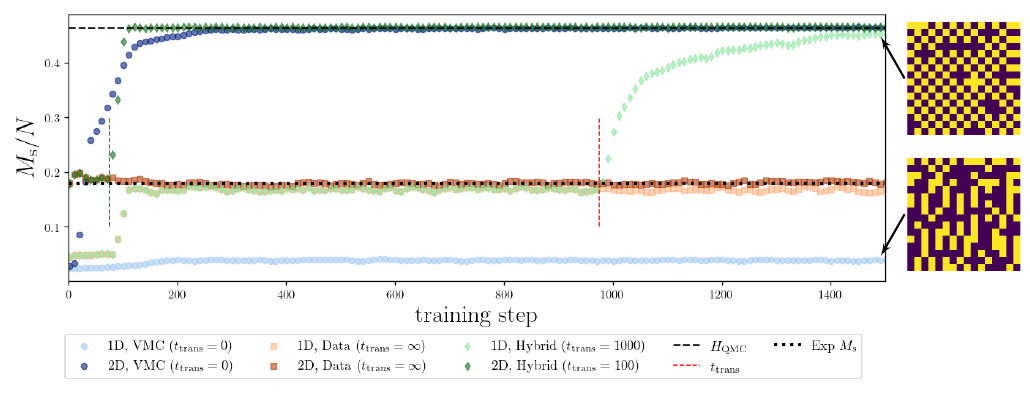}
    \vspace*{-7mm}
    \caption{
    Convergence of the estimated checkerboard order parameter $M_{\mathrm{s}}$ for 1D and 2D RNN wavefunctions trained in a purely Hamiltonian-driven setting (blue), a purely data-driven setting (orange), and in the hybrid data- then Hamiltonian-driven setting (green). The target state is deep in the checkerboard phase ($\delta/\Omega = 3.173$), meaning it is highly ordered. The value of the order parameter given by quantum Monte Carlo is marked by the black dashed line. The value of the order parameter calculated from the raw measurement data is marked by the black dotted line. For hybrid training, the transition from data-driven to Hamiltonian-driven training is marked by the red dashed line. On the right, samples drawn from the hybrid trained 1D RNN wavefunction (top) and variationally optimized 1D RNN wavefunction (bottom) demonstrate how the 1D RNN wavefunction is unable to capture the 2D order of the checkerboard phase without the use of the experimental data.
    }
    \label{fig:stagmagtraining}
\end{figure*}

\begin{equation*}
    \langle\hat{\mathcal{O}}\rangle_{\text{RNN}} = \sum_{\boldsymbol\sigma} P(\boldsymbol\sigma)\frac{\langle \boldsymbol\sigma\vert \hat{\mathcal{O}} \vert \Psi_{\mathcal{W}}\rangle}{\langle \boldsymbol\sigma\vert \Psi_{\mathcal{W}}\rangle},
\end{equation*}
where $P(\boldsymbol\sigma) = \frac{\vert\langle \Psi_{\mathcal{W}}\vert\boldsymbol\sigma\rangle\vert^2}{\sum_{\boldsymbol\sigma\prime} \vert\langle \Psi_{\mathcal{W}}\vert\boldsymbol\sigma\prime \rangle \vert^2}$. Instead of defining $\hat{\mathcal{O}}_{\text{loc}}$, as done for the energy calculation, we will consider what happens when the off-diagonal operator $\hat{\mathcal{O}}$ operates on the bra $\langle \boldsymbol\sigma\vert$. If our purely off-diagonal operator $\hat{\mathcal{O}}$ maps a given configuration $\boldsymbol\sigma$ to some other configuration $\boldsymbol\sigma\prime$ multiplied by a constant $C_{\mathcal{O}}$, that is
$\hat{\mathcal{O}}\vert\boldsymbol\sigma\rangle = \text{C}_{\mathcal{O}} \vert \boldsymbol\sigma\prime\rangle$, we can rewrite our expectaion value of $\hat{\mathcal{O}}$ as
\begin{equation*}
    \langle\hat{\mathcal{O}}\rangle_{\text{RNN}} = \sum_{\boldsymbol\sigma} \text{C}_{\mathcal{O}} P(\boldsymbol\sigma)\frac{ \Psi_{\mathcal{W}}(\boldsymbol\sigma\prime)}{\Psi_{\mathcal{W}}(\boldsymbol\sigma)},
\end{equation*}
where we have used the fact that $\Psi_{\mathcal{W}}(\boldsymbol\sigma) = \langle \boldsymbol\sigma\vert\Psi_{\mathcal{W}}\rangle$. As done previously, we can estimate this expectation value with a sample average
\begin{equation*}
    \langle\hat{\mathcal{O}}\rangle_{\text{RNN}} \approx \frac{1}{N_s}\sum_{\boldsymbol\sigma\sim p_{\text{RNN}}(\boldsymbol\sigma;\mathcal{W})} \text{C}_{\mathcal{O}} \frac{ \Psi_{\mathcal{W}}(\boldsymbol\sigma\prime)}{\Psi_{\mathcal{W}}(\boldsymbol\sigma)},
\end{equation*}
where we have approximated $P(\boldsymbol\sigma)$ with the frequency with which $\boldsymbol\sigma$ is sampled from the RNN wavefunction.

To make this calculation more concrete, consider the off-diagonal operator $\hat{\mathcal{O}} = \hat{\sigma}^x_i$. In this case, $\text{C}_{\mathcal{O}} = 1$, and a configuration $\boldsymbol\sigma$ is mapped to a different configuration $\boldsymbol\sigma_i\prime$, where $(\boldsymbol\sigma_i\prime)_j = \delta_{i,j} - \boldsymbol\sigma_j$ (which is to say $\boldsymbol\sigma_i\prime$ is identical to $\boldsymbol\sigma$ with only the i-th spin flipped). By this definition, the spatially averaged expectation value of $\hat\sigma^x_i$ can then be written as,
\begin{equation*}
    \langle\hat{\sigma}^x\rangle_{\text{RNN}} = \frac{1}{N_s}\sum_{\boldsymbol\sigma\sim p_{\text{RNN}}(\boldsymbol\sigma;\mathcal{W})}  \frac{1}{N}\sum_{\boldsymbol\sigma_i\prime} \frac{ \Psi_{\mathcal{W}}(\boldsymbol\sigma_i\prime)}{\Psi_{\mathcal{W}}(\boldsymbol\sigma)}.
\end{equation*}
There are N terms in the sum over $\boldsymbol\sigma_i\prime$, where N is the number of spins, as there are N-many ways to flip one spin in each sample.
It is worth noting that, in addition to efficient sampling, direct access to the wavefunction amplitudes of specific samples (e.g. $\Psi_{\mathcal{W}}(\boldsymbol\sigma_i\prime)$) is essential for estimating off-diagonal operators. Autoregressive generative models provide explicit access to the learned approximation of the Born distribution, allowing these amplitudes to be directly and efficiently calculated.

\section{What is Learned from Measurement Data}\label{app:stagmagtraining}

In an attempt to understand why data-enhancement provides a dramatic improvement in performance for our 1D RNN wavefunctions, we tracked the checkerboard order parameter throughout the training of our models. Based on the observation that the hybrid-trained 1D RNNs provide accurate estimates for the order parameter deep in the checkerboard phase ($\delta/\Omega > 2.5$), we hypothesized that it was easier for the model to learn about this two-dimensional ordering from the shots of projective measurements than from the Hamiltonian. This hypothesis is supported by Fig.~\ref{fig:stagmagtraining}, which shows how estimates of this order parameter evolve over the course of the training for a ground state deep in the checkerboard phase (i.e. highly ordered).

When trained purely variationally, the 1D RNN is clearly unable to learn the two-dimensional ordering of the system. The estimated order parameter remains close to zero for the entirety of the training. The inability of the model to correctly learn the ordering of the phase is further exemplified by the bottom configuration displayed on the left side of Fig.~\ref{fig:stagmagtraining}, which is a sample drawn from the model at the end of its purely variational optimization. When trained using only the experimental measurement data, the 1D RNN is able to capture some non-zero ordering, but the value learned by the model is limited by the data. Even the 2D RNN wavefunction, when trained on data alone, is not able to produce an estimate for $M_s$ that is notably better than the value of the order parameter calculated directly from the raw experimental data. More concretely, both orange curves are limited by the black dotted line. When trained using the hybrid data- then Hamiltonian- driven method, however, the 1D RNN is able to accurately capture the value of the order parameter corresponding to this ground state. In fact, it is in close agreement with the value obtained from our quantum Monte Carlo simulations after $\sim 1400$ training steps. The top configuration displayed on the left side of Fig.~\ref{fig:stagmagtraining} is a sample drawn from this hybrid-trained model and clearly contains much stronger ordering. Interestingly, for the 1D RNN, only combining these two types of training allows this model to capture the ordered phase. 
The 2D RNN wavefunctions are able to capture the ordering of this phase, likely due to their ability to pass information in more than one direction, but the data-enhancement still provides a speedup in the time to convergence. In this case, the pre-training again seems to provide more information about the order of the phase, as seen by comparing the values of the order parameter early in the training for the 2D RNN trained purely variationally and using the hybrid method. 

These dynamics can be understood by realizing that our data-enhancement provides the model with exemplary configurations of the target ground state or states close to the ground state early on in the training and optimizes the model to produce samples \emph{like} the ones provided. This serves to better initialize the model and, because the model is able to produce samples similar to the data used during pre-training, the state space that must be traversed during the subsequent variational training is reduced. This is only the case when the data is a good representation of the ground state being targeted during the Hamiltonian-driven training. 

We have thus demonstrated how certain features of a phase or, more specifically, a ground state in that phase can be more easily learned from measurement data than directly from the Hamiltonian during variational optimization. This result is foundational in our belief that this hybrid training approach will prove useful for learning ground states in exotic phases of matter. For instance, it has been shown that the non-trivial sign structure of some ground states are difficult to learn~\cite{westerhout2019neural,bukov2021learning}. Perhaps even a limited amount of data in a few bases would provide valuable information in learning such a ground state.

\section{Experimental Parameters and RNN Wavefunction Hyperparameters}\label{app:params}

We focus on measurement data spanning the disorder-to-checkerboard phase transition on a Rydberg atom array~\cite{ebadi2021quantum}. The parameters of the experiments that produced this data are summarized in Table~\ref{tab:expparams}. 
Throughout this work we employ recurrent neural networks as parameterized wavefunctions. Table~\ref{tab:rnnparams} summarizes the relevant hyperparameters for the models we used. It is important to note that the data-driven training steps listed in this table were used for the simulations reported in the main text. In Appendix~\ref{app:data}, the optimal number of data-driven training steps was used for each simulation. In these cases, $t_{\text{trans}}$ was automatically taken to be the training epoch at which the minimum energy was achieved during the data-driven training.

\vspace*{1.5mm}

\begin{table}[h!]
    \centering
    \begin{tabular}{|m{0.4\linewidth}|m{0.4\linewidth}|}
    
        \hline
         & \\
        Lattice Size &  $16\times16$ atoms\\
         & \\
        \hline
         & \\
        Sweep Rate & 15 MHz/$\mu$\\
         & \\
        \hline
         & \\
        Rabi Frequency $\Omega$ & 4.24\\
          & \\
        \hline
         & \\
         Effective Lattice & 1.15\\
         Spacing $R_{\text{b}}/a$ & \\
          & \\
        \hline
         & \\
        Detuning $\delta$ values & -1.545 to 13.455 by 0.5 \\
          &  \\
        \hline
         & \\
        Critical detuning  $\delta_{\text{c}}$ &  4.76(5)\\
         & \\
        \hline
    \end{tabular}
    \caption{Experimental parameters}
    \label{tab:expparams}
\end{table}

\begin{table}[]
    \centering
    \begin{tabular}{|m{0.5\linewidth}|m{0.2\linewidth}|m{0.2\linewidth}|}
    
        \hline
         & &\\
         & 1D RNN & 2D RNN\\
         & &\\
         \hline
         & &\\
        RNN Cell &  GRU & Tensorized  \\
         & & GRU \\
         & & \\
        \hline
         & & \\
        Number of hidden units $N_h$ & 32 & 16\\
          & &\\
        \hline
         & &\\
        Total number of parameters & 3522 & 4674 \\
          & & \\
        \hline
          & & \\
         Learning Rate (Hamiltonian-& $1\times10^{-3}$ & $1\times10^{-3}$\\
         driven or data-driven only) & &\\
           & & \\
        \hline
         & & \\
         Number of data-driven & 1000 & 100\\
        training steps (during hybrid & &\\
          training) & &\\
          & & \\
        \hline
          & & \\
         Learning Rate (Hamiltonian- & $5\times10^{-5}$ & $1\times10^{-3}$\\
         driven,  during hybrid training) & &\\
          & & \\
        \hline
    \end{tabular}
    \caption{RNN wavefunction hyperparameters}
    \label{tab:rnnparams}
\end{table}

\bibliography{references}{}

\end{document}